\documentstyle[aps,preprint]{revtex} 
\begin{document} 

\draft
\title{NEW PHYSICS
POTENTIAL \\ WITH A NEUTRINO TELESCOPE}
\author{\bf{N.~Arteaga-Romero$^{a}$,
C.~Carimalo$^{a}$ \\
A.~Nicolaidis$^{b}$,
O.~Panella$^{c}$
and
G.~Tsirigoti$^{b}$}\footnote{e-mail addresses : carimalo@in2p3.fr,
nicolaid@ccf.auth.gr, o.panella@perugia.infn.it}}
\address{ 
$^{(a)}$~Laboratoire de Physique
Corpusculaire, Coll\`ege de France,\\ 11 Pl. Marcelin Berthelot F-75231,
Paris Cedex 05, France\\ 
$^{(b)}$~Department of Theoretical
Physics, University of Thessaloniki, GR-54006 Thessaloniki, Greece\\ 
$^{(c)}$~INFN, Sezione di Perugia, Via A. Pascoli, I-06123 Perugia,
Italy}
\maketitle
\begin{abstract}
Active Galactic Nuclei are considered as sources of neutrinos, with
neutrino energies extending up to 10$^{18}$ eV. It is expected that these
highly energetic cosmic neutrinos will be detected by the neutrino
telescopes, presently under construction. The detection process is very
sensitive to the total $\nu_{\mu} N$ cross-section. We examine how
$\sigma_{tot}(\nu_{\mu} N)$ changes at high energies, by the single
production of excited fermions ($\mu^{*}, \nu_{\mu}^{*}$). For parameters
(masses, couplings) of the excited fermions allowed by the experimental
constraints, we find that for energies of the incoming $\nu_{\mu}$ above
100 TeV the cross-section for single production of $\mu^{*},
\nu_{\mu}^{*}$ supersedes the standard total cross-section.
\end{abstract}

%
%

  High energy is the prerequisite for the study of matter in
shorter distances, or the production of new massive states. 
It appears that the most powerful accelerators are the cosmic
accelerators in the outer space. Active Galactic Nuclei
(hereafter AGN) are the most powerful radiation sources known in
the Universe and they have long been considered as prodigious
particle accelerators and beam
dumps\cite{Robson,kazanas,Gaisser}. As the term
indicates, AGN are the central regions of certain galaxies in
which the emission of radiation can rival or even surpass the
total power output of the entire galaxy by as much as a thousand
fold. All this power is emitted from a region which is extremely
small by galactic standards. Typical AGN luminosities are in the
range $10^{44}\hspace{0.25em} \hbox{erg/s}$ to
$10^{47}\hspace{0.25em} \hbox{erg/s}$. The
tremendous power output suggests that the source that powers AGN
is gravity, i.e. matter accretion into a supermassive black hole
located at the center of the galaxy. 
Within AGN, particles and in
particular protons can be accelerated to very high energies. The
maximum energy $E_{max}$ attained by protons is determined
by balancing the proton acceleration rate with the proton energy
loss rate. Simple estimates indicate that $E_{max} \sim
10^{18}$  eV. The relativistic protons may interact with matter
or radiation in the AGN to produce pions whose decay products
include photons and neutrinos. It is expected that neutrinos and
photons are equally abundant and their spectra should be in
general of the same form as the parent proton spectrum. AGN
photons have been already observed by the EGRET instrument
\cite{Fich} aboard the Compton Gamma-Ray Observatory and the
Whipple Observatory \cite{Punch}. While photons are subject to
photon-photon absorption within the volume of AGN (whose optical
depth is proportional to the photon energy, thus cutting off the
highest energy photons), neutrinos suffer no such absorption,
indicating that AGN could be the most luminous high energy
neutrino sources in the Universe. The diffuse isotropic neutrino
flux from all AGN has been estimated \cite{Stecker,Nellen} and
it might be observable by the neutrino telescopes, presently
under construction.

Cosmic high energy muon neutrinos can be observed by detecting
the long range muons produced in charged current muon
neutrino-nucleon interactions. The effective detector volume is
enhanced in proportion to the range of the produced muon
(typically several kilometers). At high energies (above 1 TeV)
the produced muon is almost aligned to the parent muon neutrino,
thus knowledge of the muon direction fixes the origin of the neutrino
in the sky. Neutrino telescopes \cite{Gaisser,telescope} will
provide then a unique window to our cosmos. In addition to
opening new territory in astronomy, neutrino telescopes may
help, because of the enormous energies of the particle
interactions involved, to study fundamental physics in the
ultrahigh energy (UHE) regime. Specific signatures of "new
physics" at a neutrino telescope have been analyzed already.
These include the search for substructure of the elementery
particles \cite{Domokos}, multiple production of gauge bosons
\cite{Agnello,Morris}, scalar leptoquarks \cite{Bergstrom} and 
contact four-fermion interactions \cite{Morris}.
Detection of UHE cosmic neutrinos depends crucially upon the
total $\nu_{\mu}N$ cross-section, $\sigma_{tot} (E)$. To reduce
background, at the
detection site one looks for upward moving muons, induced
by neutrinos coming from the other side of the Earth. At energies
above a few TeV the Earth starts becoming opaque to neutrinos. The
neutrino propagation inside the Earth has been studied in
ref. 13. At very high energies and for a neutrino energy spectrum
which is not flat, the neutrino attenuation can be approximated
by the simple absorption formula
\begin{equation}
I(E,\tau) \simeq I_{0}(E) \exp \left[-\sigma_{tot}(E) \tau \right]
\end{equation} 
In the above expression $\tau$ is the total number of nucleons 
per unit area
encountered by the neutrino along its path through the Earth and
$I_{0}$ is the initial neutrino intensity. For a neutrino
going through the center of the Earth $\tau=\tau_{max}\simeq 
6\, {\times}\, 10^{33}$  $\hbox{cm}^{-2}$. Neutrino absorption is very
sensitive to $\sigma_{tot}(E)$. Any sort of new physics (new
interaction terms, production of new massive states) will
increase $\sigma_{tot}(E)$ with a subsequent dramatic reduction
of the neutrino intensity at the detection site.

In this Letter we address the implications of a composite
scenario as regards the detection of AGN neutrinos.  The idea that at an
energy scale $\Lambda_{\hbox{c}}$ quarks and leptons might show an
internal structure has been around for quite some time\cite{sub}. Various
models describing quarks and leptons in terms of {\it preon} bound states
have been proposed, but so far no consistent dynamical composite theory
has been found \cite{preons}.  
However a natural consequence of this scenario is the existence
of excited states of the ordinary fermions with masses at least of the
order of the compositeness scale.
  Effective couplings between the excited
and light leptons have been proposed, using weak isospin ($I_W$) and
hyper-charge ($Y$)  conservation. Within this model, it is assumed that
the lightness of the ordinary leptons could be related to some global
unbroken chiral symmetry which would produce massless bound states of
preons in the absence of weak perturbations due to $SU(2)\times U(1) $
gauge and Higgs interactions.  The large mass of the excited leptons
arises from the unknown underlying dynamics and {\it not} from the Higgs
mechanism. We restrict ourselves to one family and consider spin-$1/2$
excited states grouped in multiplets with $I_W=1/2 $ and $ Y=-1$,
\begin{equation} L= {\nu^{*}_{\mu} \choose \mu^{*}} \end{equation} which
can couple to the light left-handed multiplet \begin{equation} \ell_L
={{1-\gamma_5} \over 2} {\nu_\mu \choose \mu} \end{equation} through the
gauge fields ${\vec W}^{\mu}$ and $B^{\mu}$, the relevant interaction (of
magnetic type) being written \cite{Boudjema} in terms of two {\it new}
independent coupling constants $f$ and $f'$, as \begin{eqnarray} {\cal
L}_{int}& = & \frac{gf}{\Lambda_{\hbox{c}}} \bar{L}\sigma_{\mu\nu}
\frac{\vec\tau}{2} \ell_L \cdot \partial^\nu\vec{W}^\mu \cr & &
\phantom{xxxxx} +\frac{g'f'}{\Lambda_{\hbox{c}}} \biggl(-\frac{1}{2}
\bar{L}\sigma_{\mu\nu} \ell_L \biggr)\cdot \partial^\nu B^\mu +
\hbox{h.c.} \end{eqnarray} where ${\vec \tau}$ are the Pauli $SU(2)$
matrices, $g$ and $g'$ are the usual $SU(2)$ and $U(1)$ gauge coupling
constants, and the factor of $-1/2$ in the second term is the hyper-charge
of the $U(1)$ current. This effective Lagrangian has been widely used in
the literature to predict production cross sections and decay rates of the
excited particles at colliders\cite{Boudjema}. The extension to quarks and
strong interactions as well as to other multiplets and a detailed
discussion of the spectroscopy of the excited particles can also be found
in \cite{Boudjema}, while for a review of compositeness phenomenology we
refer to ref. \cite{pantrento}.  The effective interaction, written out in
terms of the physical gauge fields is \begin{equation} {\cal L}_{eff} =
\sum_{V=\gamma,Z,W}\frac{e}{\Lambda_{c}} C_{{{V \ell L}}} \bar{L}
\sigma^{\mu\nu} (1-\gamma_5) \ell \, \partial_\mu V_\nu + h.c. 
\end{equation} to be compared with the standard model interaction
\begin{equation} {\cal L}_{SM} = \sum_{V=\gamma,Z,W} e \bar{\ell'}
\gamma^\mu (A_{V\ell'\ell} -B_{V\ell'\ell} \gamma_5)  \ell \, V_\mu + h.c. 
\end{equation} The relevant effective couplings are \begin{eqnarray}
C_{W\nu\mu^{*}}=\frac{f}{2\sqrt{2}\sin \theta_{W}}\;\;\nonumber\\[1em]
C_{Z\nu\nu^{*}}=\frac{f\cot \theta_{W}+f'\tan \theta_{W}}{4}\;\;\\[1em]
C_{\gamma\nu\nu^{*}}=\frac{f-f'}{4}\;\;\nonumber \end{eqnarray} For the
Weinberg angle, we have used $\sin^{2}\theta_{W}=0.226$.  In the following
we study production of excited leptons in the collisions of UHE neutrinos
with nucleons \begin{eqnarray} {\nu_\mu N \to \nu^*_\mu
X}\;\nonumber\\[0.5em] {\nu_\mu N \to \mu^* X} \end{eqnarray}

For the partonic process $\nu_\mu q(\bar{q}) \to \nu^*_\mu(\mu^*) q'$ we
find \begin{eqnarray}\label{partonic} \frac{d{\hat
\sigma}}{dQ^2}&=&\frac{2\pi\alpha^2}{{\hat s}^2 {\Lambda_{\hbox{c}}}^2}\,
Q^2
 \sum_{ V,V'}\frac{1}{(Q^2+M_V^2)(Q^2+M_{ V'}^2)}\times \cr 
& &\biggl\{ D_{VV'}\biggl[ 2{\hat s }^2 - (2 { \hat s} - m_*^2)
(m_*^2+Q^2)\biggr] 
\pm E_{VV'} m_*^2(2{\hat s} -m_*^2-Q^2)\biggr\}\cr
& & \phantom{xxxxxxxxxxxxxxxxxxxxxxxxxxxxxxx}
\end{eqnarray}
where $m^{*}$ is the mass of the produced excited lepton and the
$\pm$ sign depends on whether the neutrino scatters off
a quark or an antiquark. The sum over $V$ restricts only to $W$
for the charged current process but includes both $\gamma$ and $Z$
for the neutral current process (see ref.~\cite{hagiwara} for
calculations of excited lepton production in $e^+e^-$ and $e p$ 
colliders).
We have also  defined
\begin{eqnarray}
D_{VV'}= 4 C_{V\nu\nu^*(\mu^*)} C_{V'\nu\nu^*(\mu^*)}
(A_{Vqq'}A_{V'qq'}+B_{Vqq'}B_{V'qq'})\;\nonumber\\[0.5em]
E_{VV'}= 4 C_{V\nu\nu^*(\mu^*)} C_{V'\nu\nu^*(\mu^*)}
(A_{Vqq'}B_{V'qq'}+B_{Vqq'}A_{V'qq'})
\end{eqnarray}
The hadronic cross-section is related to the partonic one 
by the usual convolution with the parton distribution
functions
\begin{equation}
\frac{d\sigma}{dx dQ^2}~(\nu_\mu N \to \nu^* (\mu^*) X) = 
\sum_{q} \frac{d\hat{\sigma}}
{dQ^2}~(\nu_\mu q \to \nu^* (\mu^*) q') f_q(x,Q^2)
\end{equation}

A cutoff $Q_{0}$ is introduced in the $Q^{2}$ integration to avoid the
region in which perturbative QCD is not valid.\footnote[1]{For 
$f\neq f'$ and very small
values of $Q^{2}$, coherent and incoherent elastic processes should 
contribute, thus enhancing
further the cross-section. We do not consider, in this first
explorative work, these contributions.} 
The integrated cross section is given by
\begin{equation} \sigma = \int_{(m_*^2+Q_0^2)/S}^{1} \, dx
\int_{Q_0^2}^{\hat{s} -m_*^2} dQ^2 \frac{d\sigma}{dx dQ^2} \end{equation}
where $\hat{s}= xS$ and $S=2 M_N E_\nu $ as usual. 
The parameters
$f/\Lambda_{c}$, $f'/\Lambda_{c}$ and $m^{*}$ are already
constrained by unsuccesful searches for excited leptons at colliders
and accelerators\cite{cons}.  More severely constrained are the parameters
referring to the first family of excited leptons $e^{*}, \nu_{e}^{*}$. 
As regards 
the parameters corresponding to $\mu^{*},
\nu_{\mu}^{*}$, for the purpose of numerical calculations we have chosen
two illustrative sets : 
\smallskip
\begin{displaymath} 
(i)\hspace{1em} \frac{f}{\Lambda_{c}} =
\frac{f'}{\Lambda_{c}} =0.03\, \, \hbox{GeV}^{-1} \hspace{2em} m_{*} = 130
\, \hbox{GeV}. \end{displaymath}
\begin{displaymath} 
(ii)\hspace{1em} f=0
\hspace{1em} \frac{f'}{\Lambda_{c}} = 0.03\, \, \hbox{GeV}^{-1}
\hspace{2em} m_{*} = 130 \, \hbox{GeV}.  \end{displaymath} 
\smallskip 
Both sets of parameters are the upper bounds suggested by the experimental
information\cite{cons}. For the parton distribution functions of an
isoscalar nucleon we used the GRV parametrization\cite{plothow}. Fig.1
shows $\sigma_{sm}$, the total  $\nu_\mu N$ cross-section as
given by the
standard model (solid line). In the energy range considered the standard
model prediction is relatively safe. At higher energy, the nucleon is
probed at very small $x$ values, where BFKL physics\cite{levin} might be
operative. The precise rise with the energy of $\sigma_{sm}$ has been
studied recently \cite{quigg}. In the same fig.1, we show the extra
contribution to $\sigma_{tot}(\nu_\mu N)$, if excited leptons are
produced.  In the case $f=f'$ the excited leptons $\mu^*, \nu^*_\mu$ are
produced via W and Z exchange (dotted line). In the case $f=0$, 
there is no
transition coupling to W and only the $\nu_{\mu}^{*}$ is produced, via Z
and $\gamma$ exchange (dashed line). Whenever there is a significant
coupling of $\nu_\mu$ to $\nu^*_\mu$ via a photon exchange, 
i.e. $f \not=f'$ (see
eq. 7), the photon propagator dominates over the corresponding propagators
of massive gauge bosons at low momentum transfer, with a resulting
enhanced cross-section near threshold. We observe that, for 
$f=f'$ and  with the assumed
values for the couplings and the masses of the excited leptons, the cross
section for the production of excited leptons supersedes the standard
cross-section for $\nu_\mu$ energies around 100 TeV.

At the energies
considered ($E_{\nu} > 10$ TeV) all phenomenological models
\cite{Stecker,Nellen} indicate that the diffuse isotropic neutrino flux
from all AGN dominates over the flux of atmospheric neutrinos. Therefore
any anisotropy of the measured neutrino flux at the detection site should
be attributed to the neutrino attenuation inside the Earth. The shadowing
of UHE neutrinos by the Earth involves the weak charged current,
resulting into absorption, and the weak neutral current, which provides a
redistribution of the neutrino energy \cite{Nikolaidis}. For our purposes,
we use the rough estimate provided by eq.(1). For neutrinos scratching the
Earth, $\tau$ is very small and we have access to the initial neutrino
intensity $I_{0}(E)$. For neutrinos incident at a different angle and
traversing the Earth, we obtain information about $I(E,\tau)$. Therefore
the ratio $I(E,\tau)/I_{0}(E)$ is experimentally accessible. The knowledge
of the density of the Earth through the seismic data \cite{D_A} allows
one to 
extract  $\sigma_{tot}(E)$ at high energy from the absorption factor.
Fig.2
shows the absorption factor for $\tau=2\times 10^{33} $ $\hbox{cm}^{-2}$
with $\sigma_{tot}=\sigma_{sm}$ and with
$\sigma_{tot}=\sigma_{sm}+\sigma_{new}$, where $\sigma_{new}$ originates
from the production of excited leptons.

In summary we analyzed the possibility of unravelling new physics in the
collisions of UHE cosmic neutrinos with nucleons.  The energy domain
reached is higher than the present HERA energy region and the muonic
sector (rather than the electronic) is explored. We focused our attention
unto the neutrino absorption factor, which is sensitive to the total
$\nu_{\mu} N$ cross-section. In order to  specify further the origin of
the new physics, specific signatures are needed. In our case, multiple
muons or
electromagnetic showers will indicate the production of $\mu^*$,
$\nu_{\mu}^*$. Work along these lines is in progress. 

\acknowledgments
This work was partially supported by the EU program ``Human Capital
and Mobility'', under contract No. CHRX-CT94-0450.
The authors wish to acknowledge useful discussions with 
the members of the Theory group of the Laboratoire de Physique
Corpusculaire at Coll\`ege de France.


\begin{figure}

\caption{
$\nu_\mu N $ cross-sections 
for the standard model (solid line), excited lepton
production with $f=0,f'=1$ (dashed line), with
$f=f'=1$ (dotted line).}
\end{figure}
\begin{figure}
\caption
{Absorption of the neutrino flux for $\tau=2\times 10^{33}$
$\hbox{cm}^{-2}$ 
within the standard model
(solid line) compared with the larger absorption in the
presence of the new physics effect:  excited lepton
production with $f=0, f'=1$ (dashed line) and with
$f=f'=1$ (dotted line).}
\end{figure}
\end{document}